\documentclass{mic2017}

\usepackage{mathptmx}       
\usepackage{helvet}         
\usepackage{courier}        
\usepackage{type1cm}        
\usepackage{makeidx}         
\usepackage{graphicx}        
\usepackage{multicol}        
\usepackage[bottom]{footmisc}

\makeindex             

\usepackage{graphicx}

\usepackage{amssymb}
\usepackage[misc]{ifsym} 

\usepackage{tabularx} 

\usepackage{verbatim}
\usepackage{array}
\usepackage{color}

\makeatletter
\newif\if@restonecol
\makeatother

\usepackage[ruled,slide]{algorithm2e} 

\usepackage{psfrag}    

\usepackage{subfigure} 

\usepackage{stfloats}  
\usepackage{eqparbox}

\usepackage{color} 

\usepackage{url}  
\usepackage{ifthen}  
\usepackage{multicol}   
\usepackage[utf8]{inputenc} 
\usepackage{mathtools}   %
\usepackage{arydshln}
\usepackage{bm,amsmath}

\usepackage{tikz}
\usetikzlibrary{matrix,calc}

\usepackage{amsmath}
\usepackage{eqnarray}

\begin{document}

\title{Minimum Labelling bi-Connectivity}
\author{Jos\'{e} Andr\'{e}s Moreno P\'{e}rez\inst{1} \and Sergio Consoli\inst{2}}
\institute{
  Department of Computing Engineering, IUDR, Universidad de La Laguna, Spain,
  \email{\textit{jamoreno@ull.edu.es}}
  \and
  Philips Research, High Tech Campus 34, Eindhoven 5656 AE, The Netherlands,
  \email{\textit{sergio.consoli@philips.com}}
}

\id{id}
\maketitle

\section{Introduction and problem formulation}
\label{sec-introduction}

In labelling optimization problems, 
a graph is said to be labelled if each edge has assigned a label from a finite set of labels; several edges can have the same label.
These problems consist in selecting the optimal set of labels optimizing some objective and subject to some constraints.
Several labelling problems have been proposed in the literature.
The two most studied labelling problems are the Minimum Labelling Spanning Tree Problem (MLST) \cite{MLSTP-intell-ASOC} and the Colorful Traveling Salesman Problem (CTSP) \cite{Silberholz2013}, that consist, respectively, in finding the spanning tree, and the hamiltonian cycle, that uses the minimum set of labels of the input graph.
%
Other labeling problems have been derived from the MLST problem by relaxing the connectivity property \cite{Hassin2007a}.
%
Here we propose to deal with the concept of bi-connectivity instead of that of connectivity.
The \emph{Minimum Labelling Bi-connectivity Problem} (MLBP) consists of finding the minimum size set of labels that provides bi-connectivity.
Given an undirected graph $G$, 
we say that it is connected if and only if any pair of vertices are connected or joined by some path.
An undirected graph is (single) connected if and only if for any pair of vertices there is, at least, a path joining them.
Then a graph $G$ is bi-connected if and only if there are, at least, two disjoint paths joining any pair of vertices.
The idea of disjunction among paths can be taken by considering the paths as a set of edges
or also by considering the paths as a set of vertices.
Then we say that graph $G$ is edge bi-connected if and only if for any pair of vertices there are, at least, two paths without a common edge.
Analogously, we say that graph $G$ is vertex bi-connected if and only if for any pair of vertices there are, at least, two paths without a common vertex, excluding the two joined vertices.
Bi-connected graphs or networks have important applications in robustness and resilience of transport, communication and social networks.
This condition guarantees that such networks remain connected in the event of a node or edge failure.
The classical design of communication networks have been usually based on rings.
This is because the shortest bi-connected network joining a set of vertices is a cycle or a ring passing throws all of them; i.e. a hamiltonian cycle. Therefore the 
CTSP can be used to find the minimum size set of labels
of the simplest bi-connected subgraph including all the vertices.
However, this is not equivalent to get the minimum size set of labels that provides bi-connectivity since there exist bi-connected graphs without any Hamiltonian cycle.


In the following we give some important notions. The concept of
cut-vertex is related with vertex bi-connectivity,
and the concept of
cut-edge is related with edge bi-connectivity.
A cut-vertex or articulation point is a vertex whose removal disconnects the graph.
A cut-edge or bridge is an edge whose removal disconnects the graph.
The vertex bi-connectivity is equivalent to the nonexistence of any articulation point (cut-vertex)
and the edge bi-connectivity is equivalent to the nonexistence of any bridge (cut-edge).
The connected components of an undirected graphs are its maximal connected subgraphs.
Analogously, the maximal bi-connected subgraphs of a graph are its bi-connected components, usually called \emph{blocks}.
To distinguish among blocks corresponding to edge bi-connectivity or to vertex bi-connectivity we refer to them as
edge-blocks or vertex-blocks, respectively.

Some useful conclusions derived directly from these definitions are the following.
An edge or vertex cannot belong to more than one edge-block.
Therefore edge-blocks are disjoint as set of vertices and as set of edges.
Vertex-blocks are also disjoint as set of edges but a cut-vertex belongs to more than one vertex-block.
Thus vertex-blocks are not necessarily disjoint in terms of sets of vertices.
Since two paths without a common vertex cannot have a common edge, any vertex-block is included in a edge-block.
Therefore every edge-block consists of vertex-blocks joined by cut-vertices.
Each bridge joins two vertices of different edge-blocks, and therefore of two different vertex-blocks.
Every edge that is not a bridge joints two vertices of the same vertex-block, and also of the same edge-block.
Therefore, every bridge is not in any vertex-block, neither in any edge-block.
The extremes of any bridge with degree greater than one are cut-vertices.
Finally, the set of vertex-blocks plus the set of bridges constitute a partition of the whole set of edges.
Since an edge-block consists of vertex-blocks, also the set of edge-blocks plus the set of bridges constitute another partition of the whole set of edges.
Any bridge is assumed to be a single vertex-block, then the set of vertex-blocks constitutes a partition of the set of edges.
The set of edge-blocks also constitutes a partition of the set of vertices.
However, the set of vertex-blocks does not constitute a partition of the set of vertices because a cut-vertex is in two different vertex-blocks.
Therefore a vertex-block is a set of edges and an edge-block is a set of vertices.
A single vertex is a vertex-block if and only if it is an isolated vertex
or it is the extreme of one or more edges that are bridges.

We now formulate the MLBP in a formal way. Consider a labelled, undirected graph $G$ = $(V,E,\cal{L})$, i.e. an undirected graph $G$ where every edge $e\in E$ has an unique label $l(e)\in \cal{L}$, where $V$, $E$, and $\cal{L}$ are, respectively, the sets of vertices, edges and labels.
The goal of the MLBP consists of selecting the smallest set of labels such that the corresponding subgraph is bi-connected, i.e. find the set $L \subseteq \cal{L}$ that minimizes $|L|$, where $L \subseteq \cal{L}$ and $G(L) = (V,E(L))$ is bi-connected, where $E(L) = \{ e\in E \textrm{ s.t. } l(e) \in L \}$.
Since the bi-connected maximal subgraphs are called blocks, this problem is also referred to as Minimum Labelling Spanning Block (MLSB) problem.
The MLSB problem consists of finding the spanning block $B$ having the minimum number of distinct labels.
This formulation is valid for both 
edge bi-connectivity and the vertex bi-connectivity; in these cases the problem is called, respectively, Minimum Labeling Edge-Block (MLEB) problem and Minimum Labeling Vertex-Block (MLVB).


\section{Exact solution approach}
\label{sec_algortithms}



We adapt the exact approach for the MLST problem in \cite{MLSTP-intell-ASOC} to solve both the MLEB and MLVB 
problems.
The procedure works in the space of sets of labels searching for the minimum size feasible solution.
The method is based on an $A^*$ or backtracking procedure to test the subsets of $\cal{L}$.
The algorithm performs a branch and prune procedure in the partial solution space based on a
recursive procedure that attempts to find an smaller solution from the current solution, both for edge and vertex connectivity.
The main program of the exact method calls the recursive procedure with an empty set of labels,
and iteratively stores the smallest feasible solution to date, say $L^*$.
The key procedure of the exact method is a subroutine that determines the number of blocks of the new graph $G(L)$,
for a given set of labels $L\subseteq \cal{L}$.
The recursive procedure tries to decrease the number of blocks by adding a new label to the partial solution $L$.
If $G(L)$ is (edge/vertex) bi-connected and $|L| < |L^*|$ then it becomes the smallest feasible solution; i.e. $L^* \leftarrow L$.
In the other side, if $|L| = |L^*|-1 $ and $G(L)$ is not (edge/vertex) bi-connected then
the solutions that includes $L$ are discarded and the tree is pruned.
The number of sets tested, and therefore the running time, can be shortened by pruning the search tree using this and other simple rules.

A Depth First Search (DFS) given in \cite{HT73} determines, in addition to the connected components,
the bridges and the articulation vertices of an undirected graph.
Given the bridges and articulation points we determine the edge-blocks and vertex-blocks.
DFS is a graph traversal algorithms 
that uses the edges to traverse the graph and visit its vertices.
Each time a vertex $v$ is visited, the outgoing edges $[v,w]$ are included in a list $C$ of candidate edges to be traversed.
When an edge $[v,w]$ is included in the list it is oriented from $v$ to $w$.
The algorithm starts with an empty list $C$ and visits an initial vertex.
At each step, the algorithm selects an edge $[v,w]$ from $C$ to be traversed.
If $w$ is an unvisited then edge $[v,w]$ is traversed and $w$ visited.
Otherwise, $[v,w]$ is a back edge and not traversed.
The DFS traversal algorithm stops when $C$ becomes empty.
The traversed edges constituted a tree rooted at the starting vertex.
DFS selects the edges from $C$ in LIFO (Last-In-First-Out) order
and manage list $C$ as a stack.

DFS is linear in the number of edges and is the basis for many graph-related algorithms, including topological sorts, planarity, and connectivity testing.
It may be used to determine if a graph is connected and, in the negative case, to obtain its connected components.
For this purpose, start the traversal iteratively from a non visited vertex, until all the vertices are then visited.
The vertices of each connected component are those visited in each run of the algorithm.
The number of connected components is the number of runs.
DFS was extended in \cite{Tarjan74} to determine if a graph is bi-connected and, in the negative case, to obtain its blocks.
This DFS algorithm finds the cut-vertices and bridges of the graph as follows.
The algorithm traverse the vertices and edges of the graph using the depth-first strategy
and determining a so-called directed DFS-tree.
The algorithm uses the depth of every vertex and
the lowest depth of neighbors of all its descendants in the DFS tree, its lowpoint.
The root of the DFS tree is a cut-vertex if it has more than one outgoing tree edges.
A vertex $v$, which is not the root, is a cut-vertex if it has a child $w$ such that no
back edge starting in the subtree of $w$ reaches an ancestor of $v$.
They are determined during the execution of the DFS by using the order in which the vertices are visited.
The DFS algorithm recursively get the first visited node that is reachable from $v$
by using a directed path having at most one back edge; i.e., the last edge of the path.
If this vertex was not visited before $v$ by DFS, then $v$ is a cut-vertex.
The lowpoint of $v$ can be computed after visiting all descendants of $v$ as the minimum of the depth of $v$, the depth of all neighbors of $v$
and the lowpoint of all children of $v$ in the depth-first-search tree.
Moreover, DFS traverse the edges of each vertex-block consecutively.
Therefore, the vertex-blocks are obtained using a stack to keep track of the edges being traversed.
A bridge is a back edge that constitute a vertex-blocks.
The edge-blocks are obtained by joining the blocks joined by cut-vertices.

\section{Summary and Outlook} \label{Conclusions}

We considered the Minimum Label Bi-connectivity problem (MLBP) consisting of finding the minimum set of
labels that provide bi-connectivity among all the vertices of a graph.
The graph is bi-connected if there are at least two disjoint paths joining every pair of vertices.
We considered both bi-connectivity concepts: the edge bi-connectivity where these paths cannot have a common edge
and the vertex bi-connectivity where the paths cannot have a common vertex.
This is 
a NP-hard graph combinatorial optimization problem deriving from the well-known MLST problem.
The smallest bi-connected subgraph including all the vertices is a hamiltonian cycle and the CTSP
consists of finding the minimum set of labels of a hamiltonian cycle.
However the MLBP is not equivalent to the CTSP since there are bi-connected graphs without hamiltonian cycle.
To test if a given set of labels includes a hamiltonian cycle
is NP-hard, however a DFS algorithm determines the bi-connectivity in linear time.

After providing the main definitions and formulation of the MLBP,
we scratched the details of an exact solution approach under current development.
The algorithm will be able to provide the first known gold standard solutions for the problem.
Although the method would able to get exact solutions only for relatively small
problem instances, it will be useful for the development of heuristics with high performance guarantees.
It is worth noting that the problem may be generalized to the concept of $k$-connectivity, with $k>2$.
The extension of these results and the exploration of efficient solution approaches for the generalization problem of $k$-connectivity, for $k \ge 2$, may be objects worth of future investigation.




\bibliographystyle{plain}

\end{document}